\documentclass[aps,twocolumn,prl,superscriptaddress,showpacs]{revtex4}
\usepackage{bm}
\usepackage{graphicx,amsmath,latexsym,amssymb}
\usepackage{color}

\begin{document}

\definecolor{olive}{rgb}{0,1,.4}
\title{Mesoscopic phase statistics of diffuse ultrasound in dynamic matter}

\author{M. L. Cowan}
\altaffiliation[Current address: ]
                     {Department of Physics, University of Toronto, Toronto, ON, Canada M5S 3E3}
\affiliation{Department of Physics and Astronomy, University of
Manitoba, Winnipeg, Manitoba, Canada}

\author{D. Anache-M\'{e}nier}
\affiliation{Laboratoire de Physique et Mod\'{e}lisation des
Milieux Condens\'{e}s, CNRS /Universit\'{e} Joseph Fourier, BP
166, F-38042 Grenoble Cedex 9, France}

\author{W. K. Hildebrand}
\affiliation{Department of Physics and Astronomy, University of
Manitoba, Winnipeg, Manitoba, Canada}

\author{J. H. Page}
\affiliation{Department of Physics and Astronomy, University of
Manitoba, Winnipeg, Manitoba, Canada}

\author{B. A. van Tiggelen}
\affiliation{Laboratoire de Physique et Mod\'{e}lisation des
Milieux Condens\'{e}s, CNRS /Universit\'{e} Joseph Fourier, BP
166, F-38042 Grenoble Cedex 9, France}

\date{\today}

\begin{abstract}
Temporal fluctuations in the phase of waves transmitted through a
dynamic, strongly scattering, mesoscopic sample are investigated
using ultrasonic waves, and compared with theoretical predictions
based on circular Gaussian statistics. The fundamental role of
phase in Diffusing Acoustic Wave Spectroscopy is revealed, and
phase statistics are also shown to provide a sensitive and
accurate way to probe scatterer motions at both short and long
time scales.
\end{abstract}


\maketitle

For all waves, phase is irrefutably the most fundamental property.
On macroscopic scales, however, phase is randomized by multiple
scattering or obscured by decoherence. It is now generally
accepted that a mesoscopic regime exists where wave phenomena
persist  on even hydrodynamic scales. Mesoscopic fluctuations can
sometimes be long-range and non-Gaussian \cite{rev,book}. The
universal conductance fluctuations are best known, originally
discovered for electrons  \cite{electrons}, and later also
observed with visible light \cite{lightC3} and microwaves
\cite{microwaveC3}. In the optics of soft condensed matter, the
existence of dynamic mesoscopic fluctuations has led to a new
technique called \emph{diffusing wave spectroscopy} (DWS)
\cite{dws}. In the acoustic counterpart, \emph{diffusing acoustic
wave spectroscopy} (DAWS) \cite{daws}, the fluctuations of the
scattered wave field are measured directly to probe the dynamics
of disordered media. In seismology, the closely related technique
of Coda Wave Interferometry \cite{CWI} is extending the range of
applications being studied.

For acoustic, seismic or radio waves, the phase can be easily
extracted.  While many applications, including interferometric
techniques such as InSAR \cite{insar}, make use of phase for
precise measurements, the phase of multiply scattered waves has
often been neglected, since it is generally more challenging to
extract useful information from phase in multiple-scattering
systems.  
Mesoscopic studies have revealed the fundamental relation of phase
to the screening of zeros of random fields \cite{zero}, but most
of the literature has focussed on quantities such as the
probability distribution functions of intensity, transmission or
conductance \cite{rev}, and does not address the phase directly.
Recent studies with microwaves \cite{genackphase}, infrared light
\cite{lagendijkphase} and Terahertz radiation \cite{thz} have
explored frequency correlations of the phase. In this Letter, we
study \emph{time-dependent} phase fluctuations of ultrasound in a
dynamic, strongly scattering medium, and examine the statistics of
both the wrapped and cumulative phase evolution. This combination
of theory and experiment reveals a deeper insight into the
mesoscopic physics of multiply scattered waves, and explicitly
shows the relationship between the average phase evolution of a
typical multiple scattering path - a crucial concept in D(A)WS
modelling \cite{dws,daws} - and the measured phase evolution of
the transmitted waves.  We also find that phase statistics can
sometimes provide a more accurate method of measuring the dynamics
than the field autocorrelation method that is used in D(A)WS.  In
our materials, the temporal phase fluctuations are too complex for
traditional Doppler ultrasound analysis. This Letter may be viewed
as a way of overcoming these complications.

The ability of ultrasonic piezoelectric transducers to detect the
wave field allows the phase of the scattered ultrasound to be
measured directly.  In our experiments, we used a pulsed
technique, so that the phase of multiply scattered waves along
paths spanning a narrow range of path lengths could be
investigated. For most of the experiments, the sample was a
12.2-mm-thick fluidized bed, containing 1-mm-diameter glass
spheres suspended at a volume fraction of 40\% by an upward
flowing solution of 60\% glycerol and 40\% water.  A miniature
hydrophone was used to capture the field transmitted through the
sample in a single near-field speckle spot.  The input pulses had
a central frequency of $2.25 $ MHz ($\lambda = 0.71 $ mm), were
roughly 5 periods wide, and were repeated every $2\, $ms.  Since
the beads were in constant motion, the scattered signal was
different for each input pulse, allowing the phase to be measured
as a function of the evolution time $T$ of the sample and the
propagation time  $t$ of the waves. Since the sample hardly
changed during the propagation time, the system appeared
``frozen'' to the individual pulses. At a fixed lapse time after
each pulse input pulse, a short segment (about 4.5 periods) of the
transmitted waveform was recorded. Using a simple numerical
technique \cite{cargese}, the wrapped phase $\Phi(t) \in
(-\pi:\pi]$ and the amplitude $A(t)
> 0$ in each segment were determined as a function of time from
the digitized field data. This technique is equivalent to taking a
Hilbert transform to produce the complex analytic signal $A(t)
\exp[\mathrm i(\omega t + \Phi(t))]$, where $\omega$ is the
central frequency of the pulse. To achieve good statistical
accuracy, 10 sets of 8300 consecutive pulses were recorded.

In order to gain insight into the temporal phase fluctuations of
the multiply scattered waves, we examine the statistics of the
phase evolution and its derivatives with time.  The wrapped phase
probability distribution $P(\Phi)$, which gives the probability of
measuring a phase $\Phi$ at acoustic propagation time $t$ and
evolution time $T$, was found experimentally to be constant within
statistical error, consistent with a complex random wave field
described by Circular Gaussian Statistics (CGS) \cite{goodman}. We
have extended the theory of the phase within CGS
\cite{genackphase} to deal with the statistics of phase
\emph{evolution}, which involves the change in phase, or phase
shift, with time. The joint probability distribution of $N$
complex acoustic fields recorded at evolution times $T_i$ of the
sample is,
\begin{equation}\label{Goodman}
P(\psi_{T_1}\cdots \psi_{T_N})=\frac{1}{\pi^N\det
\textbf{C}}\exp\left[ - \sum_{i,j}^N \psi_{T_i}^*
\textbf{C}_{ij}^{-1} \psi_{T_j}  \right]
\end{equation}
\noindent where $\textbf{C}_{ij}=\langle \psi_{T_i}^{}
\psi_{T_j}^*\rangle $ is the covariance matrix \cite{goodman}.  It
is convenient to use normalized fields so that
$\textbf{C}_{ii}=1$. Then, the off-diagonal elements of
$\textbf{C}_{ij}$ are equal to the field autocorrelation function
used in DAWS \cite{daws}: $\textbf{C}_{i\neq j}= g_1(T_i-T_j)$.
For $N=2$, two wave amplitudes and one phase can be integrated out
from Eq.~(\ref{Goodman}) at constant phase difference
$\Delta\Phi(\tau) =  \Phi(T+\tau) - \Phi(T)$. If we \emph{rewrap}
the phase difference into the interval $(-\pi:\pi]$, we get for
the probability distribution of phase evolution
\begin{equation}\label{ProbaDelta}
P(\Delta\Phi)=\frac{1}{2\pi}\left[\frac{1-g_1^2}{1-
F^2}\right]\left[1+\frac{ F\cos^{-1}F}{\sqrt{1- F^2}} \right],
\end{equation}
where $F\equiv g_1\cos\Delta\Phi$. As the scatterers move for time
$\tau$, the acoustic fields $\psi_T$ and $\psi_{T+\tau}$
decorrelate. This process strongly affects the statistics of the
temporal phase evolution $\Delta\Phi(\tau)$, with
$P(\Delta\Phi(\tau))$ finally approaching the flat distribution
for large time differences ($g_1=0$).

The phase dynamics can be described quantitatively in terms of the
variance of the change in phase along \emph{one typical} path
taken by the waves, $\langle\Delta
\phi^2_{\mathrm{path}}(\tau)\rangle $
, which we call simply the ``path phase variance''
\cite{footnoteAveragePhi}. 
The DAWS auto-correlation function \cite{daws} is related directly
to this variance according to $g_1(\tau) \approx\exp[-\frac{1}{2}
\langle\Delta\phi^2_{\mathrm{path}}(\tau)\rangle]$. From
Eq.~(\ref{ProbaDelta}) we can establish that the path phase
variance is significantly different to the wrapped phase shift
variance $\langle \Delta\Phi^2(\tau)\rangle$. The latter is
associated with the \emph{superposition} of the waves from all
paths at the detector that, for the phase, implies a highly
nonlinear transformation. Yet, a \emph{universal} relation
$\left\langle \Delta\Phi^2\right\rangle =
f(\langle\Delta\phi^2_{\mathrm{path}}\rangle)$ is predicted, with
no  parameter that depends 
on the details of the
dynamics. We exploit this universality below to find
$\langle\Delta \phi^2_{\mathrm{path}}(\tau)\rangle $ 
directly from the wrapped phase shift variance.  Quite
surprisingly, we will see later that unwrapping the phase destroys
this universality.

The path phase variance can be related to the particle motion
according to \cite{daws}
$\langle\Delta\phi^2_\mathrm{path}(\tau)\rangle \simeq \
\frac{1}{3}n k^2\langle\Delta
r^2_{\mathrm{rel}}(\tau,\ell^*)\rangle $. Here $k$ is the wave
vector, $n$ is the average number of scattering events and
$\langle\Delta r^2_{\mathrm{rel}}(\tau,\ell^*)\rangle$ is the
relative mean square displacement of two scatterers separated by
the transport mean free path $\ell^*$ of the sound. At early times
we expect ballistic motion, $\left\langle\Delta
r^2_{\mathrm{rel}}(\tau)\right\rangle=\left\langle\Delta
V^2_{\mathrm{rel}}\right\rangle\tau^2$, and it is convenient to
write
 $\langle\Delta
\phi^2_{\mathrm{path}}(\tau)\rangle
 = \frac{1}{3}\tau^2/\tau^2_{\mathrm{DAWS}}$,
where $\tau_{\mathrm{DAWS}}=1/\sqrt{n k^2 \left\langle\Delta
V_{rel}^2 \right\rangle}$  is  the characteristic time scale
beyond which the particle motion  destroys the correlation of the
acoustic field.

At short times and small $\Delta\Phi$, Eq.~(\ref{ProbaDelta})
simplifies to  $P(\Delta\Phi) = \frac{1}{2}\langle\Delta
\phi^2_{\mathrm{path}}\rangle/[\langle\Delta
\phi^2_{\mathrm{path}}\rangle+\Delta\Phi^2]^{3/2}$, showing
directly its dependence on $\langle\Delta
\phi^2_{\mathrm{path}}(\tau)\rangle$, and hence $\langle\Delta
r^2_{\mathrm{rel}}(\tau,\ell^*)\rangle$. This expression has the
same form as the probability distribution of the phase derivative
$\Phi' $ with evolution time,
$P(\Phi')=\frac{1}{2}Q/[Q+\Phi'^2]^{3/2}$, where
$Q=\lim_{\tau\rightarrow
0}\langle\Delta\phi^2_{\mathrm{path}}(\tau)\rangle/\tau^2=
(3\tau^2_{\mathrm{DAWS}})^{-1}
$.

\begin{figure}
\includegraphics[width=9cm]{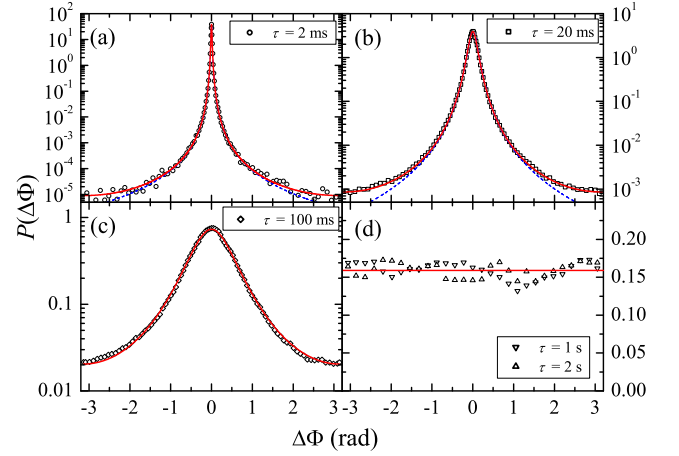}
\caption{(\textit{color online}) The observed probability
distribution of the wrapped phase evolution at five different time
intervals $\tau$ (symbols) and the corresponding theoretical
predictions (solid curves).  The dashed curves in (a) and (b) are
the small $\tau$, small $\Delta\Phi$ predictions (see text). The
only fitting parameter (via the dependence of $P(\Delta\Phi)$ on
$g_1)$) is the path phase variance
$\langle\Delta\phi_{\mathrm{path}}^2\rangle$ at time $\tau$, which
gives $\tau_{\mathrm{DAWS}} = 89$ ms. For these data, the number
of scattering events $n=34 \pm 2$. Note the wide variation in
vertical scales from (a) to (d).} \label{fig:Pwrap}
\end{figure}

Figure 1  shows our experimental data for $P(\Delta\Phi)$ at five
values of $\tau$, along with fits to Eq.~(\ref{ProbaDelta}). The
early times show a narrow peak centered at $\Delta\Phi=0$, which
broadens as the particles move further from their original
positions.  As $\tau$ gets larger, the probability distribution is
indeed seen to approach the flat distribution [Fig. 1(d)].  The
agreement between theory and experiment is excellent over the
entire range of phases and times, and for $P(\Delta\Phi)$ spanning
nearly seven orders of magnitude. The fits provide accurate
measurements of $\langle\Delta \phi^2_{\mathrm{path}}(\tau)\rangle
$ and hence of the relative mean square displacement of the
particles.  Alternatively, by using the universal relationship
$\left\langle \Delta\Phi^2\right\rangle =
f(\langle\Delta\phi^2_{\mathrm{path}}\rangle)$ (Fig. 2(a)),
$\langle\Delta
\phi^2_{\mathrm{path}}(\tau)\rangle $ can directly be determined
from the measured variance $\left\langle
\Delta\Phi^2\right\rangle$ - 
a simpler procedure than
fitting 
$P(\Delta\Phi)$. Both methods work well so long as
$\left\langle\Delta \Phi^2\right\rangle$ is less than its upper
limit of $\pi^2/3$ when the phase difference distribution has
become flat. In Fig. 2(b), $\left\langle\Delta
r^2_{\mathrm{rel}}\right\rangle$ measured from the wrapped phase
fluctuations and the conventional field autocorrelation are
compared.  The agreement between the two methods is excellent,
giving direct experimental confirmation of the universal
relationship shown by the solid curve in Fig 2(a).

In cases where the noise in the measured signals affects the
amplitude rather than the phase (eg. gain or DC offset
fluctuations), the phase method is more robust for small $\tau$.
This is illustrated in the inset to Fig. 2(b), which shows the
effect of 2$\%$ random gain fluctuations in the field data; this
amplitude noise clearly degrades the measurement of $\langle\Delta
r^2_{\mathrm{rel}}\rangle$ from $g_1$, but does not affect the
phase measurement.

\begin{figure}
\includegraphics[width=8cm]{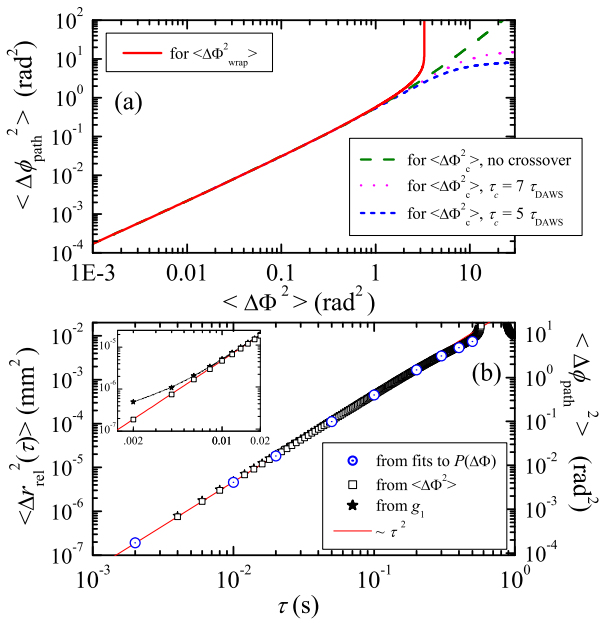}
\caption{(\textit{color online})
\definecolor{olive}{rgb}{.25,.68,.31} (a) The universal relationship
(solid curve), calculated from Eq.~\ref{ProbaDelta}, between the
variance of the phase shift along one path ($y$-axis), and the
measured phase of the transmitted field ($x$-axis). Dashed and
dotted lines apply when the phase is unwrapped, making the
relation explicitly depend on the motion of the particles. (b) The
relative mean square displacement of the particles,
$\left\langle\Delta r^2_{\mathrm{rel}}(\tau)\right\rangle $ (left
axis), determined from the wrapped phase via the corresponding
$\langle\Delta\phi_{\mathrm{path}}^2\rangle$ (right axis). We
compare the results from the wrapped phase shift distribution 
(\textcolor{blue}{${\bigodot}$}) and variance (\textcolor{olive}{$
\square  $}) with traditional DAWS measurements ($\star$). The
inset shows the effect of amplitude noise (see text). }
\label{phase2dr}
\end{figure}

\begin{figure}
\includegraphics[bb=4cm 9cm 17cm 20cm ,width=7cm]{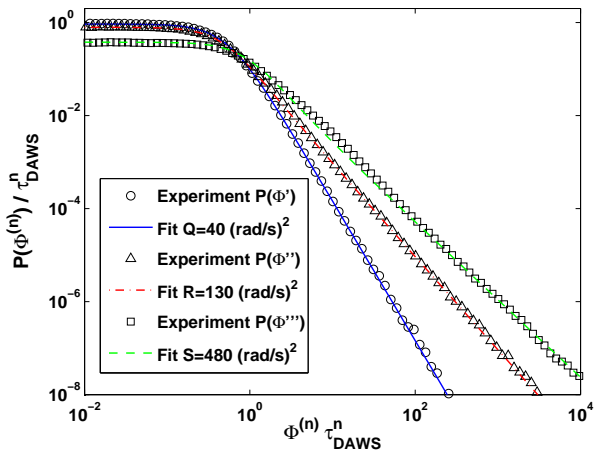}
\caption{(\textit{color online}) Comparison of theory and
experiment for the phase derivative distribution functions
$P(\Phi^{(n)})$, where $n = 1, 2$ or $3$ denotes the $n^{th}$
derivative of $\Phi$ with respect to evolution time $T$. From $Q$
= 40 rad$^2$/s$^{-2}$, we find $\tau_{\mathrm{DAWS}}$ = 91 ms.
 } \label{ProbaDeriv}
\end{figure}

By considering the joint probability distribution of $N=4$ fields
in Eq.~(\ref{Goodman}) and by integrating out one phase and four
amplitudes, we have obtained an analytic expression for the joint
probability distribution of the first three phase derivatives with
evolution time $P(\Phi',\Phi'',\Phi''')$ \cite{webpage}, from
which the individual distribution functions $P(\Phi')$,
$P(\Phi'')$ and $P(\Phi''')$ can be computed. They depend on three
parameters $Q, R$ and $S$, that in turn relate to time derivatives
of the field correlation function $g_1(\tau)$ at $\tau= 0$: $Q
=-g_1''(0)$, $R=-[{g_1^{(4)}(0)-g_1''(0)^2}]/g_1''(0) $, and $S  =
\left[{-g_1^{(6)}(0)+\left(g_1^{(4)}(0)^2/{g_1''(0)}\right)}\right]/\left[{g_1^{(4)}(0)-g_1''(0)^2}\right]$.
The fits to the three distributions give the values of $Q$, $S$
and $R$. (Fig.~\ref{ProbaDeriv}). These in turn provide a
sensitive probe of the early time behavior of the particle motion,
$\langle\Delta r^2_{\mathrm{rel}}\rangle$ in powers of
$x=\tau/\tau_{\mathrm{DAWS}}$: $\left\langle\Delta
r^2_{\mathrm{rel}}\right\rangle= 324x^2- 57x^4-3.3x^6$ $\mu
\mathrm{m}^2$. We emphasize that, by using this method, details
about the motion  \emph{up to the 6$^{th}$ power in time} can be
retrieved, which would be impossible from the conventional DAWS
method. Figure~\ref{ProbaDeriv} also shows that both theoretical
and experimental distributions follow an asymptotic power law
decay with exponents $-3$, $-2$, $-\frac{5}{3}$ (which suggests
$-(1+\frac{2}{n})$ for the $n^{th}$ derivative). These slopes
provide a \emph{fit-independent} test for CGS.

To investigate the evolution of the phase over longer times, we
study the \textit{cumulative} (unwrapped) phase $\Phi_c(T)$, which
can be obtained by adding or subtracting $2\pi$ whenever there is
a jump of $\pm 2\pi$ in the wrapped phase. The cumulative phase
can be defined as $\Phi_c(T)=\int_0^{T}\Phi'(\tilde{T}) \,
\mathrm{d}\tilde{T}$, and is, by construction, a continuous random
variable that is no longer constrained to the interval
$(-\pi,\pi]$.  Its ensemble-average vanishes for fields described
by CGS. For sufficiently long time intervals, we expect the
cumulative phase shift $\Delta \Phi_c(\tau)$ to approach the
normal distribution with zero mean \cite{patrick}. Its variance is
related to the cumulative phase derivative correlation function,
$C_{\Phi'}(\tau)\equiv\left\langle\Phi'(T-\frac{1}{2}\tau)\Phi'(T+\frac{1}{2}\tau)\right\rangle$,
which in CGS has the simple analytic form
$C_{\Phi'}(\tau)=\frac{1}{2}(\ln g_1)''\ln(1-g_1^2)$
\cite{anacheEPL}.  Fig. 4(a) compares theory and experiment for
$C_{\Phi'}$, where predictions based on a simple empirical
crossover model for the particle dynamics are also included
\cite{daws}, for which $\left\langle\Delta
r^2_{\mathrm{rel}}(\tau)\right\rangle=\left\langle\Delta
V^2_{\mathrm{rel}}\right\rangle\tau^2/\left(1+\tau^2/\tau_c^2\right)$.
The best fit is obtained for $\tau_c=7\tau_{\mathrm{DAWS}}$,
showing that both $\tau_{\mathrm{DAWS}}$ and $\tau_c$ can be
determined from $C_{\Phi'}$.

\begin{figure}
\includegraphics[width=8cm]{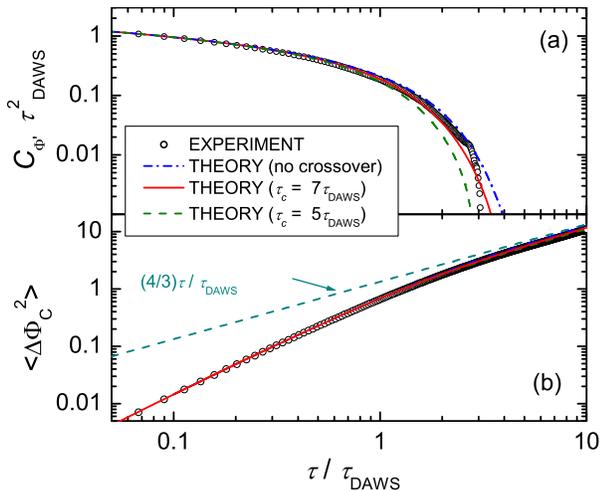}
\caption{(\textit{color online})  Comparison of theory and
experiment for (a) the phase derivative correlation function
$C_{\Phi'}$, (b) the cumulative phase shift variance
$\left\langle\Delta \Phi^2_c(\tau)\right\rangle$.
}
\label{fig:cumphase}
\end{figure}

The cumulative phase shift variance can be calculated from
$C_{\Phi'}$ since $\left\langle\Delta
\Phi^2_c(\tau)\right\rangle=2\int_0^{\tau}d\nu (\tau-\nu)
C_{\Phi'}(\nu)$ \cite{anacheEPL}. Recalling the expression for
$C_{\Phi'}(\tau)$ reveals that the variance of cumulative phase
evolution is determined by $g_1$ and its  first two derivatives.
This destroys the universal relation with the path phase variance,
but at the same time this increases the sensitivity to details in
particle motion at long times (see Fig.~\ref{phase2dr}(a)).  At
short times, the cumulative phase variance increases as a power
law with a logarithmic correction, $\left\langle\Delta
\Phi^2_c(\tau)\right\rangle =
\frac{1}{3}\tau^2/\tau_{\mathrm{DAWS}}^2[1.5-\ln(\tau/\sqrt3 \,
\tau_{\mathrm{DAWS}})]$, while at long times, the cumulative phase
shift evolves as a 1D random process with finite diffusion
constant:  $ \left\langle\Delta\Phi^2_c(\tau)\right\rangle
\rightarrow D_{\Delta\Phi}\, \tau $ (Fig.~\ref{fig:cumphase}(b)).
For $\tau_c\gg\tau_{\mathrm{DAWS}}$, we find $D_{\Delta\Phi}=
\frac{4}{3} \tau_{\mathrm{DAWS}} $. The cumulative phase variance
is quite different from the path phase variance
$\langle\Delta\phi_\mathrm{path}^2\rangle$, which has a finite
diffusion constant \emph{only} if the particles undergo Brownian
motion, which they do not here. By comparing measured and
theoretical $\left\langle \Delta\Phi_c^2(\tau)\right\rangle$ in
Fig.~\ref{fig:cumphase}(b), the accurate value of
$\tau_{\mathrm{DAWS}}= 89$ ms was deduced from an appropriate
translation along the $x$ direction.

We have studied the phase evolution of ultrasonic waves in
strongly scattering, dynamic media.  It is important to
discriminate the random phase evolution along one scattering path,
usually studied in D(A)WS, from the observed phase evolution in a
single speckle spot. Our experiments are extremely well modelled
by circular Gaussian statistics.  This theory accurately predicts
the behavior of the wrapped phase difference probability
distribution, the variance of both the wrapped and cumulative
phase shifts, and the phase derivative distributions and
correlation function. The excellent agreement of theory and
experiment has allowed us to relate the observed fluctuations in
phase evolution to the relative mean square displacement of the
scatterers.  The phase statistics are sensitive probes of the
particle motion on both short and long time scales, and can
provide  more accurate information than the more traditional field
fluctuation measurements.

We wish to thank NSERC for its support, and T. Norisuye for
assisting with some of the data analysis.


\begin{thebibliography}{99}
\bibitem{footnoteAveragePhi} Since our samples do not
undergo any uniform dilation, such as could arise from changes in
wave velocity due to temperature changes, the average phase
$\langle\Delta\phi_{\mathrm{path}}\rangle$ is zero, and the
variance $\langle\Delta \phi^2_{\mathrm{path}}\rangle $
characterizes the
dynamics. 
\end{thebibliography}
\end{document}